\documentclass[graybox,footinfo]{svmult}

\smartqed
\usepackage{mathptmx}       
\usepackage{helvet}         
\usepackage{courier}        
\usepackage{type1cm}        
\usepackage{graphicx}       

\usepackage{array,colortbl}
\usepackage{amsmath,amsfonts,amssymb,bm} 
\usepackage{mathtools}
\usepackage{subfig}
\usepackage{bbold} 

\DeclareFontFamily{U}{mathx}{\hyphenchar\font45}
\DeclareFontShape{U}{mathx}{m}{n}{
      <5> <6> <7> <8> <9> <10>
      <10.95> <12> <14.4> <17.28> <20.74> <24.88>
      mathx10
      }{}
\DeclareSymbolFont{mathx}{U}{mathx}{m}{n}
\DeclareFontSubstitution{U}{mathx}{m}{n}
\DeclareMathAccent{\widecheck}      {0}{mathx}{"71}

\newcommand{\group}[1]{\mathcal{#1}}
\newcommand{\Ugroup}[1]{\group{U}(#1)}
\newcommand{\SU}[1]{\group{SU}(#1)}
\newcommand{\dint}[1]{\int\!\dif #1\,}
\newcommand{\intIn}[2]{\int_{#2}\!\dif #1\,}
\newcommand{\dif}{\text{d}}
\DeclareMathOperator{\ex}{\mathrm{e}}
\newcommand{\e}[1]{\ex^{#1}}
\newcommand{\MC}{\text{MC}}

\newcommand{\tr}[1]{\text{tr}{#1}}
\newcommand{\trlong}[1]{\tr{\left[{{#1}}\right]}}
\newcommand{\diag}[1]{\text{diag}{#1}}
\newcommand{\order}{\mathcal{O}}

\renewcommand{\email}[1]{\emailname: #1} 

\usepackage{mathptmx}       
\usepackage{helvet}         
\usepackage{courier}        

\usepackage{makeidx}         
\usepackage{graphicx}        
\usepackage[bottom]{footmisc}

\usepackage{latexsym}
\usepackage{amsmath}
\usepackage{amsfonts}
\usepackage{amssymb}
\usepackage{bm}

\usepackage{url}
\usepackage{algorithm}
\usepackage{algorithmic}
\usepackage[misc,geometry]{ifsym}

\spdefaulttheorem{assumption}{Assumption}{\upshape \bfseries}{\itshape}
\spdefaulttheorem{algo}{Algorithm}{\upshape \bfseries}{\itshape}

\newcommand{\R}{{\mathbb{R}}} 

\DeclareSymbolFont{bbold}{U}{bbold}{m}{n}
\DeclareSymbolFontAlphabet{\mathbbold}{bbold}

\DeclareSymbolFont{bbold}{U}{bbold}{m}{n}
\DeclareSymbolFontAlphabet{\mathbbold}{bbold}

\usepackage{microtype} 

\usepackage[colorlinks=true,linkcolor=black,citecolor=black,urlcolor=black]{hyperref}
\usepackage{bookmark}
\makeatletter 
  \providecommand*{\toclevel@author}{999}
  \providecommand*{\toclevel@title}{0}
\makeatother

\begin{document}

\title*{Avoiding the sign-problem in lattice field theory}
\author{Tobias Hartung \and Karl Jansen \and Hernan Le\"ovey \and Julia Volmer}
\institute{
Karl Jansen \and Julia Volmer - Speakers,
\at DESY Zeuthen, Platanenallee 6, 15738 Zeuthen, Germany \\
\email{karl.jansen@desy.de}, \email{julia.volmer@desy.de}
\and 
Tobias Hartung
\at Department of Mathematics, Kings College London, Strand, London
WC2R 2LS, United Kingdom\\
\email{tobias.hartung@kcl.ac.uk}
\and
Hernan Le\"ovey
\at Structured Energy Management, Axpo Trading, Parkstrasse 23, 5400
Baden, Germany\\
\email{HernanEugenio.Leoevey@axpo.com}
}
\maketitle

\abstract{In lattice field theory, the interactions of elementary
  particles can be computed via high-dimensional integrals. 
  Markov-chain Monte Carlo (MCMC) methods based on importance sampling are 
normally efficient to solve most of
  these integrals. But these methods give large errors for oscillatory
integrands, exhibiting the so-called sign-problem. We developed new
quadrature rules using the symmetry of the considered systems 
to avoid the sign-problem in physical one-dimensional models for the resulting
high-dimensional integrals. This article gives a short introduction to
integrals used in lattice QCD where the interactions of gluon and
quark elementary particles are investigated, explains the alternative
integration methods we developed and shows results of
applying them to models with one physical dimension. 
The new quadrature rules avoid the sign-problem and can therefore be
used to perform simulations at until now not reachable regions in parameter
space, where the MCMC errors are too big for affordable sample sizes. However, it is still a challenge to develop these techniques further 
for applications with physical higher-dimensional systems.}


\section{Introduction}
\label{sec:intro}
\begin{figure}
\subfloat[MC sampling points]{\includegraphics[width=.45\textwidth]{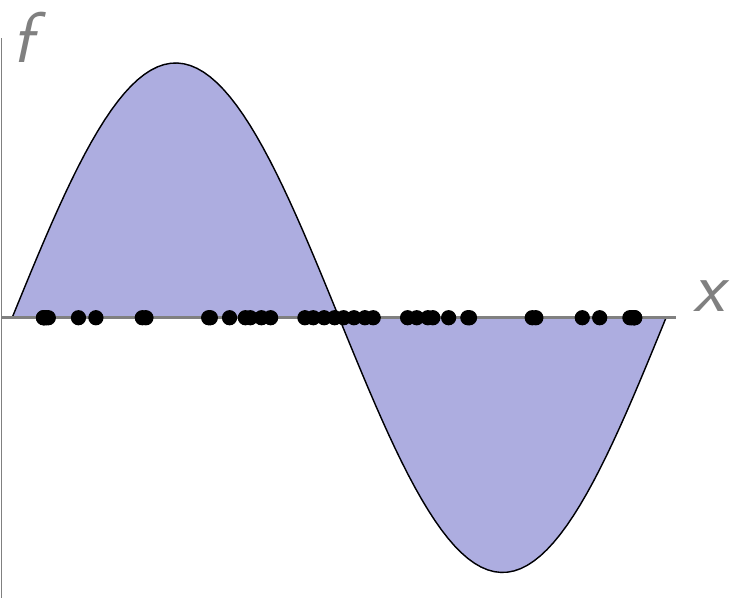}
\label{pic:intro_MC}}
\hfill
\subfloat[Symmetric sampling points]{\includegraphics[width=.45\textwidth]{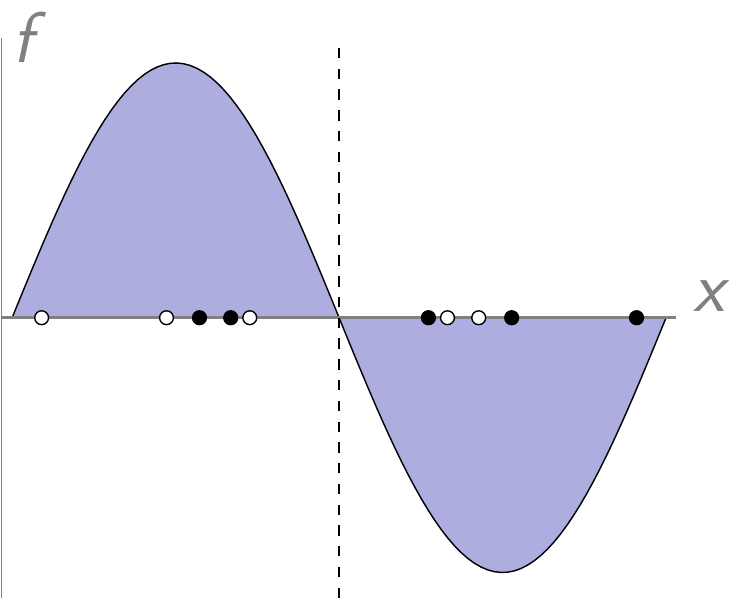}
\label{pic:intro_sym}}
\caption{MC integration of an oscillatory function results in large
  errors, known as the sign-problem. This problem is due to the
  non-cancellation of positive and negative contributions to the
  quadrature rule (a). Choosing sampling points by using the
  symmetry of the underlying model results in an exact quadrature rule
  (b).}
\end{figure}

Monte Carlo (MC) methods are in general very efficient to
solve high-dimensional integrals. They use the law of large numbers to
approximate an integral with quadrature rules that use random
sampling points. But MC methods are highly \emph{inefficient} for oscillatory
integrand functions, e.g. the function shown in Figure \ref{pic:intro_MC}. An {\em exact} integration of oscillatory functions would, 
of course, 
result in the cancellation of large negative and positive
contributions to the integral - in the example in Figure
\ref{pic:intro_MC} this would give an integral of zero. However, the random choice of sampling points
in MC methods, shown as black points in Figure \ref{pic:intro_MC}, does lead only to approximate cancellation when the number 
of points is relatively big and hence, it is very difficult to obtain 
accurate results with affordable sample sizes. 
This non-perfect cancellation of negative and positive parts in the integration
method, usually resulting in large
quadrature rule errors, is called the \emph{sign-problem}. The sign-problem is for
example the reason why important physical interactions in the early
universe cannot be simulated which could explain why there is more
matter than anti-matter in our universe today. 
To acquaint better knowledge of these
fundamental phenomena, it is essential to develop alternative
quadrature rules to MC that avoid the sign-problem. 

In physical applications, the
function to-be-integrated describes some characteristic in a given physical model.
We investigated
methods that use some symmetry of the physical model to result in the
exact cancellation of positive and negative parts in the quadrature
rule. If the model behind the function in Figure \ref{pic:intro_sym} has a
reflection symmetry, few MC sampling points - in black - can be chosen
and together with their reflected - white - points they form a set of sampling
points that results in an exact quadrature rule. In this specific
example even one MC point with its reflection point would give an exact
result, for more complicated functions more sampling points are
needed.

This article first gives a short introduction to the high-dimensional
integrals that have to be solved in particle physics, more precisely
in lattice QCD. Readers that are mostly interested in the integration
methods can easily skip this part. The main part of this article presents the
methods we developed and tested to avoid the sign-problem for high-dimensional
integration in physical one-dimensional systems.

We found that symmetrically chosen quadrature rules can avoid the
sign-problem and can efficiently be applied also to high-dimensional integrals. These
rules can help to perform simulations in important,
not-yet reachable regions in parameter space, at least in physical
one-dimensional systems so far. To
apply them to higher physical dimensions, in particular to physical four-dimensional systems 
in high energy physics as lattice QCD, they
clearly need to be developed further.

\section{Integration in lattice QCD}
\label{sec:QCD}

In theoretical physics the interaction between elemetary particles such as the electron, 
is described by {\em quantum field theories} (QFT), see e.g. \cite{Peskin:1995ev}.
The mathematical formalism in QFT defines particles as 
classical fields that are functions in three space dimensions and
one time dimension,
$P(x,y,z,t)$. Operators, $O[P]$, are functionals of these fields and describe
the interactions between them. An expectation value $A$ of this
interaction or
operator $O[P]$, also called amplitude, is computed via the path
integral,
\begin{align}
  A = \frac{\int O[P] B[P] \,\dif P }{\int B[P] \,\dif P}.
  \label{equ:lattice_amplitude}
\end{align}
$\dint{P}$ is the infinite-dimensional integration over all possible
states of the field $P$ in time and space. The path integral 
becomes a well defined expression, if a Euclidean
metric is used and the fields are defined on a finite dimensional, discrete 
lattice\footnote{For an alternative definition using the $\zeta$-regularization
see~\cite{Hartung:2018usn,Hartung:2019fcf}.}. In \eqref{equ:lattice_amplitude}, $B[P]$ is called the
Boltzmann-weight and provides a probability which weights the particle (field) 
interactions. The denominator in \eqref{equ:lattice_amplitude} insures
the proper normalization of $A$. The expectation value $A$ is
interesting because physical observables can be derived
from it and their numerical values can be compared with experimental
results or can give new results that are not yet possible to reach with experiments.

In {\em lattice field theory}, space-time and the involved functionals
$O[P]$ and $B[P]$ are discretized in Euclidean space, such that
\eqref{equ:lattice_amplitude} can be computed numerically. Often, the
Boltzmann-weight is a highly peaked function suggesting that this
computation can be done using importance sampling techniques. In most
computations, this importance sampling is done by a Markov chain MC
(MCMC) algorithm using a Markov chain that leaves the distribution
density $\frac{ B[P] }{\int B[P] \dif P} $ invariant. To compute $A$
numerically, four-dimensional space-time is discretized on a
four-dimensional lattice with four directions $\mu \in \{1,2,3,4\}$
, lattice sites
$\bm{n} \in \Lambda = \{(n_1, n_2, n_3, n_4) | n_1, n_2, n_3, n_4 \in
\{1,...,d\}\}$ and discretized fields $P$. This results
in an $4d$-dimensional integration over the Haar measure of the compact
group $\SU{3}$. For real applications, $d$ can be very large, reaching
orders of magnitude of several thousands nowadays. Thus, we are left
with an extremely high dimensional integration problem. Moreover, for
some physically very important questions MCMC methods cannot be
applied succesfully. This concerns, for example, the very early
universe or the matter anti-matter asymmetry which leads to our sheer
existence. Thus, a number of interesting questions remain completely
unanswered and it is exactly here where new high dimensional
integration methods could be extremely helpful
 
\begin{figure}
\includegraphics[width=1\textwidth]{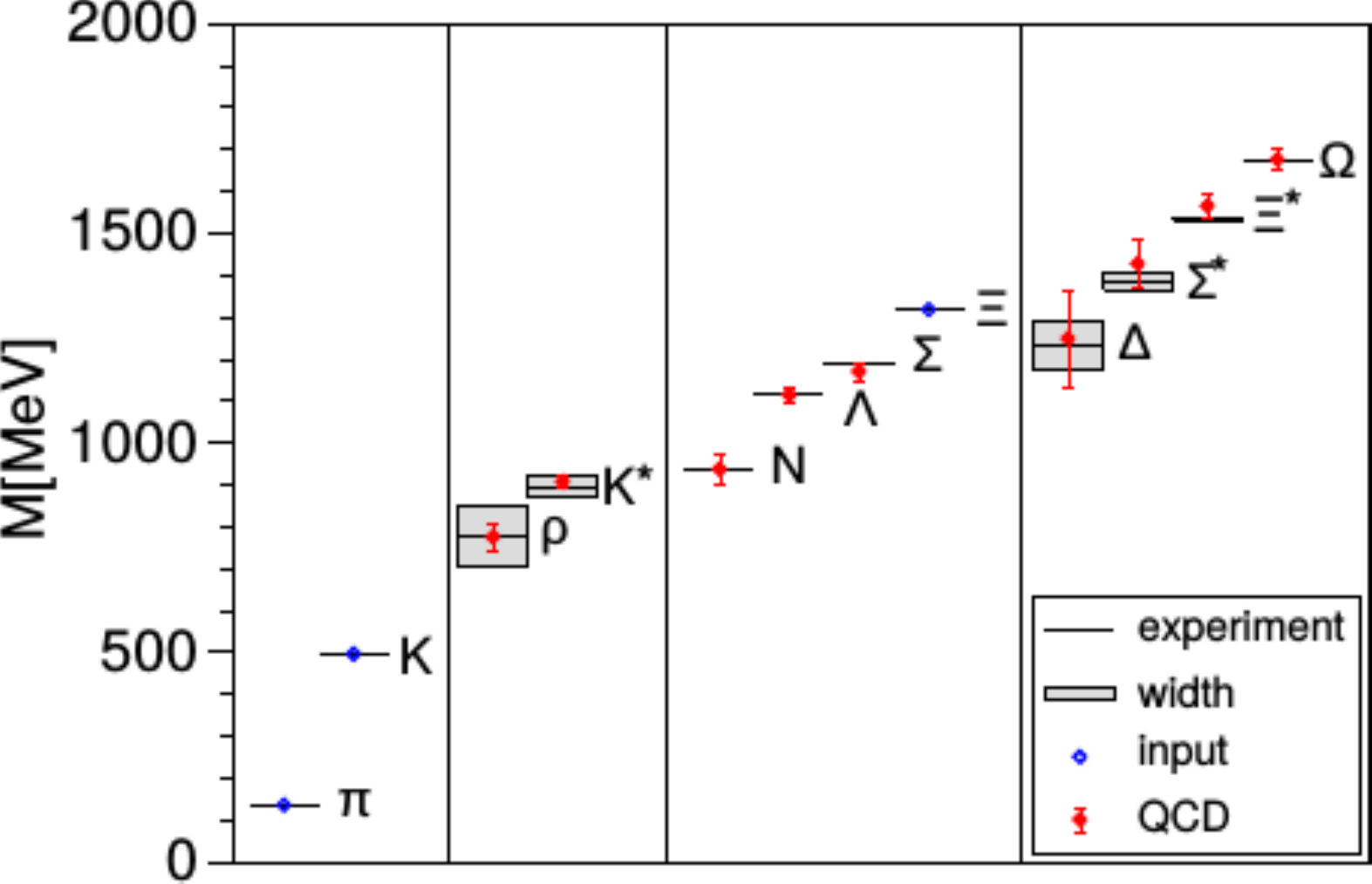}
\caption{The via lattice QCD computed masses of different composite
  particles (dots with vertical error bars) agree with the experimentally
  measured values (horizontal lines with error boxes)
  \cite{Durr:2008zz}. The masses of $\pi$, $K$ and $\sum$ (dots
  without error bars) were input values
to the computation.}
\label{pic:lattice_massSpectrum}
\end{figure}

Still, the MCM methods have led to very successful computations already. 
By performing numerical 
computations on massively parallel super computers 
a very impressive result
of such a lattice MCMC can be obtained: namely, the mass spectrum of the
lightest composite particles made out of quarks and gluons that agrees
completely with the experimental values, see Figure
\ref{pic:lattice_massSpectrum}. To get similar precise results for
other, more error-prone observables, research is going on to develop new methods to make this
high-dimensional integration faster and the results more precise.

A more detailed introduction to lattice QCD is for example given in the 
textbooks~\cite{Gattringer:2010zz,DeGrand:2006zz,Rothe:1992nt}.

\section{Quadrature rules for one-dimensional lattices}
In lattice QCD, the amplitude of interactions between quarks and gluons
in physical four-dimensional space-time can be computed via a high-dimensional
integral using the Haar-measure over the compact group $\SU{3}$, see
section \ref{sec:QCD}. This integral is typically solved numerically using MCMC
methods. If the integrand is an oscillatory function, this method
results in the sign-problem that gives large errors and avoids
physical insights in important processes. We developed alternative methods that
avoid the sign-problem and at the same time are efficient for high-dimensional
integration over compact groups. Due to various complications with
physical four-dimensional lattice QCD, we developed and tested the methods for
physical one-dimensional models that involve low-dimensional and
high-dimensional integration over compact groups.
As suggested in section \ref{sec:intro}, we
developed quadrature rules using the symmetry of the models.

This section is structured from low-dimensional to
high-dimensional integration: First, it introduces symmetric quadrature rules for
one-dimensional integration over compact groups to avoid the
sign-problem here. Then, it presents the
recursive numerical integration (RNI), a
method to reduce high-dimensional integrals to nested one-dimensional
integrals. Finally, it shows how to combine both methods to avoid the
sign-problem for
high-dimensional integration over compact groups. For all three presented
methods, the section shows results of applying them to simple
physical, one-dimensional
models. More detailed explanations of the methods and
applications can be found in \cite{Volmer2018New}.

\subsection{Avoiding the sign-problem in physical one-dimensional systems}
\label{ssec:sym0}
The sign-problem can already arise in a one-dimensional integration, solving 
\begin{align}
I(f) = \int_{G} f(U) \, \dif U
  \label{equ:sym0_int}
\end{align}
with MC methods over the Haar-measure of
$G \in \{\Ugroup{N}, \SU{N}\}$. Finding an alternative suitable
quadrature rule $Q(f)$ ad-hoc to approximate this integral is not
straightforward. The articles \cite{Bloch:2013ara, Bloch:2013qva}
suggest that using symmetrically distributed sampling points can be beneficial
for avoiding the sign-problem, possibly resulting in an exact
cancellation of positive and negative contributions to the integral,
as stated in section \ref{sec:intro}. The article of Genz \cite{genz}
gives efficient quadrature rules for integrations over spheres,
choosing the sampling points symmetrically on the spheres. We searched
for measure preserving homeomorphisms to apply the symmetric quadrature rules on spheres to
the integration over compact groups. This section describes the two
steps to create the symmetric quadrature rules $Q(f)$ for
\eqref{equ:sym0_int}:
\begin{enumerate}
  \item[Sym 1.] Rewrite the integral $I(f)$ over the compact group $G$
    into an
    integral over spheres. We restricted ourselves to $G \in \{\Ugroup{1}, \Ugroup{2}, \Ugroup{3}, \SU{2}, \SU{3}\}$.
  \item[Sym 2.] Approximate each integral over one spheres by a
    symmetric quadrature rule as proposed in
    Genz \cite{genz}, and combine
    them to a product rule $Q(f)$.
\end{enumerate}
Finally, this section shows results of applying $Q(f)$ to the
one-dimensional QCD model with a sign-problem. A more detailed
explanation of the method can be found in \cite{Ammon:2016jap, Ammon:2016ztz}.

By finding measure preserving homeomorphisms between the compact groups and products of
spheres we created polynomially exact quadrature rules for compact groups. The
application of these rules to the one-dimensional QCD model gave
results on machine precision where the standard MC method
shows a sign-problem. Therefore the symmetric quadrature rules avoid
the sign-problem and give rise to solve integrals in beforehand
non-reachable parameter regions.

\subsubsection{Sym 1. Rewriting the integral}
\label{sssec:sym0_connection}
The symmetric quadrature rules of Genz \cite{genz} are designed for the
integration over $k$-dimensional spheres $S^k$. To use them for the
integration over the compact groups $\Ugroup{N}$ and $\SU{N}$ with
$N\in\{2,3\}$ in \eqref{equ:sym0_int}, the compact
groups have to be associated with spheres. The facts that $\Ugroup{N}$ is
isomorphic to the semidirect product of $\SU{N}$ acting on
$\Ugroup{1}$ $\big(U(N) \cong SU(N) \rtimes U(1)\big)$, that $\Ugroup{1}$ is
isomorphic to $S^1$ $\big(\Ugroup{1} \cong S^1\big)$ and that $\SU{N}$ is a
principal $\SU{N-1}$ bundle over $S^{2N-1}$ result in
\begin{align}
  \SU{N} &\simeq S^3 \times S^5 \times ... \times S^{2N-1},\\
  \Ugroup{N} &\simeq S^1 \times S^3 \times ... \times S^{2N-1}.
\end{align} 
Then, the
integral over the Haar-measure of $G$ in \eqref{equ:sym0_int} can be rewritten as the integral
over products of spheres,
\begin{align}
      \intIn{U}{G} f(U) = &
                            \int_{S^{2N-1}} \Bigg(
                            \int_{S^{2N-3}} \bigg( \cdots
                            \int_{S^{n+2}} \Big(
                            \int_{S^{n}} \nonumber \\
  & \qquad \qquad f\big(\Phi(\bm{x}_{S^{2N-1}}, \bm{x}_{S^{2N-3}}, \ldots,
    \bm{x}_{S^{n+2}},\bm{x}_{S^{n}} )\big)\\
  & \qquad \qquad \qquad \qquad \qquad \qquad \dif \bm{x}_{S^{n}}
                            \Big) \dif \bm{x}_{S^{n+2}} \cdots
                            \Big) \dif \bm{x}_{S^{2N-3}}
                            \Big) \dif \bm{x}_{S^{2N-1}}, \nonumber
  \label{equ:sym01_integralOverSpheresToCompactGroups}
\end{align}
with $n=1$ for $\Ugroup{N}$ and $n=3$ for $\SU{N}$
\cite{Ammon:2016zur}. Here, $\bm{x}_{S^k}$ is an element on the
$k$-sphere and $\Phi:\ \bigtimes_j S^{2j-1} \rightarrow
G$ with $G \in \{\Ugroup{N}, \SU{N}\}$ is a measure
preserving homeomorphism. We found the homeomorphisms $\Phi_G \equiv \Phi$ for the compact groups $G \in
\{\Ugroup{1}, \Ugroup{2}, \Ugroup{3}, \SU{2}, \SU{3}\}$:

\begin{itemize}
\item For $\SU{2}$, $\Phi$ is an isomorphism, given by
\begin{align}
  \Phi_{\SU{2}}: S^3 &\rightarrow \SU{2},\nonumber\\
  \bm{x} &\mapsto   
  \begin{pmatrix} 
    x_1 + ix_2 & -(x_3 + ix_4)^* \\ 
    x_3 + i x_4 & \phantom{-}(x_1 + ix_2)^* 
  \end{pmatrix}.
\end{align}

\item For $\SU{3}$, spherical coordinates of $S^5$ are needed,
\begin{align}
  \Psi: [0,2\pi)^3 \times [0,\frac{\pi}{2}) &\rightarrow S^5,
  \nonumber\\
  (\alpha_1, \alpha_2, \alpha_3, \phi_1, \phi_2) &\mapsto 
  \begin{pmatrix}
  \cos\alpha_1 \sin\phi_1\\
  \sin\alpha_1 \sin\phi_1\\
  \sin\alpha_2 \cos\phi_2 \sin\phi_2\\
  \cos\alpha_2 \cos\phi_2 \sin\phi_2\\
  \sin\alpha_3 \cos\phi_1 \cos\phi_2\\
  \cos\alpha_3 \cos\phi_1 \cos\phi_2
  \end{pmatrix}.
  \label{equ:sym0d_trafoS5}
\end{align}
Then, $\Phi$ is given by
\begin{align}
  \Phi_{\SU{3}}: S^5_1 \times S^3 &\rightarrow \SU{3},\nonumber\\
  (\bm{x}, \bm{y}) &\mapsto A(\Psi^{-1}(\bm{x})) \cdot
                     B(\bm{y}),
\end{align}
with the matrices
\begin{align}
  A(\Psi^{-1}(\bm{x})) &= \scalebox{.7}{\mbox{\ensuremath{\displaystyle
  \begin{pmatrix}
    \e{i\alpha_1} \cos\phi_1 & 0 & \e{i\alpha_1} \sin\phi_1\\
    -\e{i\alpha_2} \sin\phi_1\sin\phi_2 & \e{-i(\alpha_1+\alpha_3)}
    \cos\phi_2 & \e{i\alpha_2} \cos\phi_1\sin\phi_2 \\
    -\e{i\alpha_3} \sin\phi_1\cos\phi_2 & -\e{-i(\alpha_1+\alpha_2)}
    \sin\phi_2 & \e{i\alpha_3} \cos\phi_1\cos\phi_2
  \end{pmatrix}}}},
  \label{equ:sym0d_A}\\
  B(\bm{y}) &= 
  \begin{pmatrix}
    x_1 + ix_2 & -(x_3 + ix_4)^* & 0\\ 
    x_3 + i x_4 & \phantom{-}(x_1 + ix_2)^* & 0\\
    0 & 0 & 1
  \end{pmatrix}.
  \label{equ:sym0d_B}
\end{align}
$\Psi^{-1}(\bm{x})$ is the inverse transformation of
\eqref{equ:sym0d_trafoS5} from Euclidean to spherical coordinates.
$S^5_1$ denotes $S^5$ without its poles, $\phi_1=0$ or
$\phi_2=0$, because at these points the inverse transformation is not unique.
The therefore excluded set is a null set, thus $\Phi_{\SU{3}}$ can still be
used in \eqref{equ:sym01_integralOverSpheresToCompactGroups}.

\item For $\Ugroup{1}$, $\Phi$ is an isomorphism,
\begin{align}
  \Phi_{\Ugroup{1}}: S^1 &\rightarrow \Ugroup{1},\nonumber\\
  \alpha &\mapsto \e{i\alpha},
\end{align}
with $\alpha \in [0, 2\pi)$.

\item For $\Ugroup{2}$, $\Phi$ is an isormophism,
\begin{align}
  \Phi_{\Ugroup{2}}: S^3 \times S^1 &\rightarrow \Ugroup{2},\nonumber\\
  (\bm{x}, \alpha) &\mapsto \Phi_{\SU{2}}(\bm{x}) \cdot \diag(\e{i\alpha},1).
\end{align}

\item For $\Ugroup{3}$, $\Phi$ is given by
\begin{align}
  \Phi_{\SU{3}}: S^5_1 \times S^3 \times S^1 &\rightarrow \Ugroup{3},\nonumber\\
  (\bm{x}, \bm{y}, \alpha) &\mapsto \Phi_{\SU{3}}(\bm{x}, \bm{y}) \cdot
                             \diag(\e{i\alpha},1,1).  
\end{align}

\end{itemize}

\subsubsection{Sym 2. Quadrature rule for spheres}
\label{sssec:sym0d_integrationOverSpheres}
With the measure preserving homeomorphism $\Phi$ in section
\ref{sssec:sym0_connection}, the integral \eqref{equ:sym0_int} can be
written as an integral over a product of spheres as in
\eqref{equ:sym01_integralOverSpheresToCompactGroups}. To approximate
the full integral numerically, one can use a product quadrature rule with
quadratures $Q_{S^k}(g)$ that are specifically designed for
integrations over spheres. The full integral can be computed
efficiently if the number of involved spheres is small. As pointed out
in the last subsection, in practice we are interested to build product rules for at most $ S^5_1 \times S^3 \times S^1 $. The quadratures over each sphere can be built in many ways. Since we are aiming for resulting quadratures that exhibit some symmetry characteristics to hopefully overcome the sign-problem, it seems that
quadrature rules given in \cite{genz} exhibit all requiered properties, i.e. high accuracy due to polynomial exactness over spheres, numerical stability of the resulting weights, and beeing fully symmetric. The quadratures over each sphere take the form
\begin{align}
  Q_{S^k}(g) = \sum_{\gamma=1}^{N_{\text{sym}}} w_\gamma \,\, g(
                    \bm{t}_\gamma).
  \label{equ:sym0_quadratureRuleForOneSphere}
\end{align}
The sampling points $\bm{t} \in S^k$ are chosen symmetrically on the
$k$-sphere and are weighted via $w \in \R$. The specific definitions
of $\bm{t}$, $w$ and $N_{\text{sym}}$ for
different $k$ are given in \cite{genz}. (Note that in this
  reference, the notation $U_k$ is equivalent to the here used $S^{k-1}$.) It is
possible to randomize these quadrature rules, such that an error estimate
for each quadrature rule can be computed via independent replication \cite{genz}.

The final quadrature rule $Q(f)$ of the full integral in
\eqref{equ:sym01_integralOverSpheresToCompactGroups} is a combination
of different single-sphere quadrature rules given in
\eqref{equ:sym0_quadratureRuleForOneSphere}.
Due to the symmetric choice of the sampling points on spheres, the rule
$Q(f)$ is
in the following called \emph{symmetrized quadrature rule}. A more
detailed description of $Q_{S^k}(g)$ and $Q(f)$ is given in
\cite{Volmer2018New}, section 6.1.

\subsubsection{Application to one-dimensional QCD}
\label{sssec:1dQCD}
We applied these constructed quadrature rules to physical
one-dimensional QCD problems \cite{Bilic:1988rw}, which is a
simplified model of strong interactions in elementary particle
physics. This model is a good test model because it can be solved
analytically, giving a well defined measure for the uncertainties
computed by different numerical integration methods. This model has
one integration variable $U \in G$ and three real input parameters: a mass $m$, a chemical potential
$\mu$ and a length scale $d$. A small mass ($m \ll d\mu$) introduces a
sign-problem which makes it very hard for standard methods as MC to compute
amplitudes as in \eqref{equ:lattice_amplitude} numerically.

We computed the chiral condensate in this model, given by
\begin{align}
  \chi = \frac{\int_{G} \partial_m B[U] \, \dif U}
  {\int_{G} B[U] \,\dif U},
  \label{equ:sym0d_chiralCondensate}
\end{align}
with the Boltzmann-weight
\begin{align}
  B[U] = \det \left(c_1(m) + c_2(d,\mu) U^\dagger + c_3(d,\mu) U \right),
  \label{equ:sym0d_partitionFunctionWithDeterminant}
\end{align}
expressed via the parameters
\begin{align}
  c_1(m) &= \prod_{j=1}^L \tilde{m}_j,
  &\tilde{m}_1 &= m, \nonumber\\
  & &\tilde{m}_j &= m + \frac{1}{4 \tilde{m}_{j-1}}
                \quad \forall j \in \{2, 3, ..., d-1\}, \nonumber\\
  & &\tilde{m}_d &= m + \frac{1}{4 \tilde{m}_{d-1}} + \sum_{j=1}^{d-1}
                \frac{(-1)^{j+1} 2^{-2j}}{\tilde{m}_j 
                \prod_{k=1}^{j-1} \tilde{m}_k^2}, \\
  c_2(d, \mu) &= 2^{-d} \e{-d \mu}, &&\\
  c_3(d, \mu) &= (-1)^d 2^{-d} \e{d\mu}.&&
\end{align}
For brevity, the dependencies of these parameters are in the following only written
when needed.

In all numerical calculations, we first computed both numerator and
denominator of \eqref{equ:sym0d_chiralCondensate} separately and then
divided them. We computed the numerator by symbolically
differentiating $B[U]$ and computing the integral over the result
numerically.

We compared the results for $\chi$ using the symmetrized quadrature
rules that are
described in \ref{sssec:sym0d_integrationOverSpheres}, with a standard
integration method, ordinary MC sampling. The latter quadrature rule
is given by
\begin{align}
  Q(f) = \frac{1}{N_\MC} \sum_{\gamma=1}^{N_{\text{MC}}} \,\, f(V_\gamma),
  \label{equ:sym0d_partitionFunctionMC}
\end{align}
where the $V$ are matrices that are chosen randomly from a uniform
distribution.
We chose $N_{\MC}$ to be as large as the number of used symmetric sampling
points.

Because the analytic results of $\chi$ can be calculated
straightforwardly, we computed the error estimates of the numerical
solutions - MC and
symmetrized quadrature rules - directly via the relative deviation from the analytic value,
\begin{align}
  \Delta \chi = \frac{|\chi_{\text{numerical}} -
  \chi_{\text{analytic}}|}{|\chi_\text{analytic}|}
  \label{equ:sym0d_errorInPlots}
\end{align}
and derived the standard deviation of this
error by repeatedly using on the one hand the MC quadrature rules with
different random matrices $V$'s and on the other hand the randomized symmetrized
quadrature rules as indicated in section \ref{sssec:sym0d_integrationOverSpheres}.

\begin{figure}[t]
  \includegraphics[width=1.\textwidth]{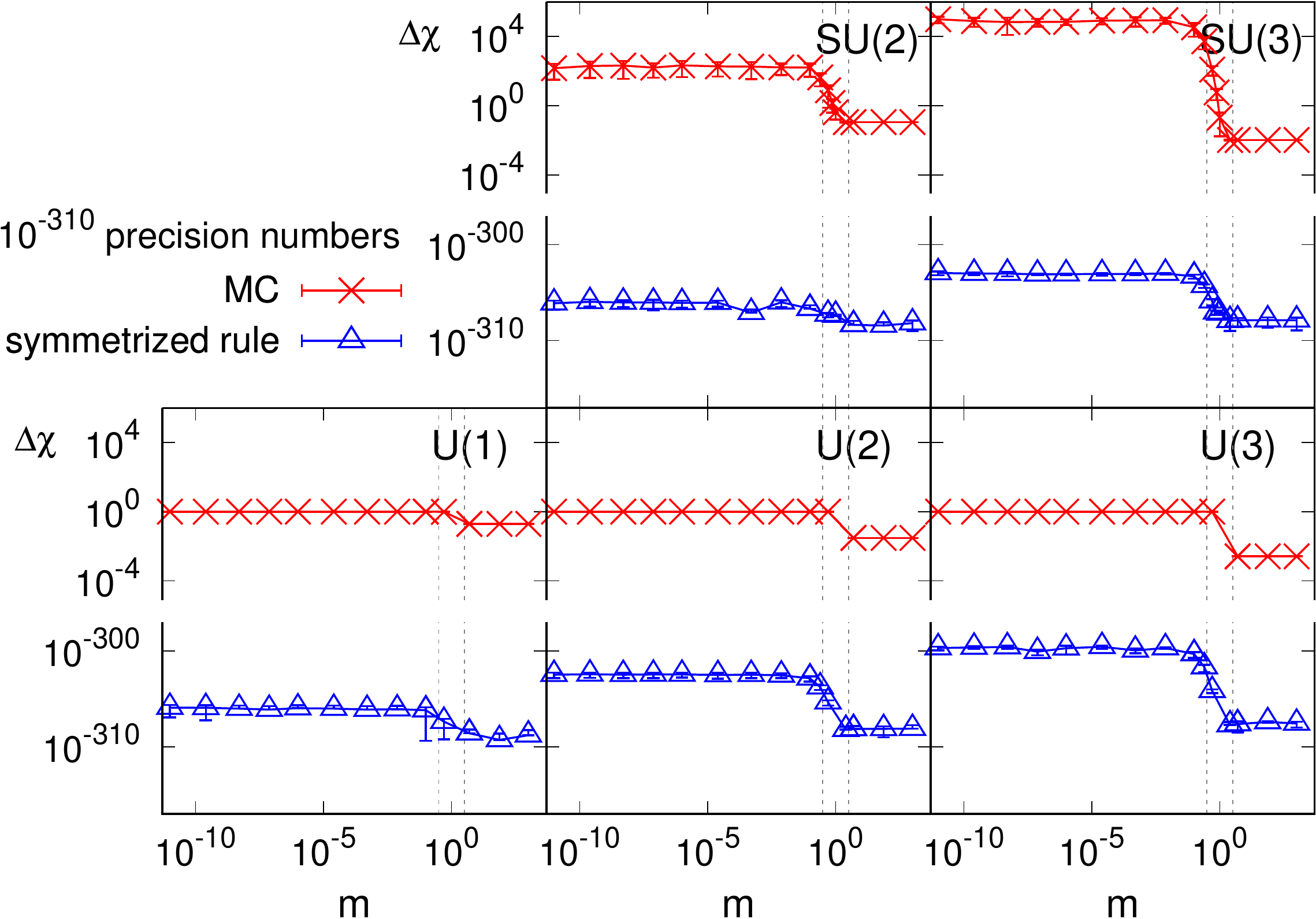}
  \caption{The sign-problem arises for MC results with small $m$
    constants, giving errors of the order of one. On the contrast, the
    symmetrized quadrature rules avoid the sign-problem in this region
    completely, giving errors approximately at machine precision for
    all shown groups.}
  \label{pic:sym0d_chiralCondensate_error_1024bitPrecision}
\end{figure}

The results for $\Delta \chi$ of both MC and symmetrized quadrature rule
can be roughly
split into a small $m$ ($m<10^{-1}$), a large $m$ ($m > 10^{0.5}$) and a
transition region, shown in Figure \ref{pic:sym0d_chiralCondensate_error_1024bitPrecision} for constant $\mu=1$ and
$d=8$, extended 1024-bit machine precision and different compact
groups. For both quadrature rules we used the sampling sizes $N \equiv
N_{\text{sym}} = N_\MC = 8$ for $\SU{2}$, $N = 96$ for $\SU{3}$, $N=4$
for $\Ugroup{1}$, $N=32$ for $\Ugroup{2}$ and $N=384$ for $\Ugroup{3}$.

First, we describe the MC results:
In the small $m$ region, $\Delta \chi$ for all groups are large - equal
or larger than one. It can be shown that in this region the numerator
of $\chi$ is such small that the MC evaluation cannot resolve these
values for affordable sample sizes, resulting in large errors \cite{Volmer2018New}.
This is the manifestation of the sign-problem, making it almost
impossible to compute reasonable values of $\chi$ with MC in the small
$m$ region.
On the other side, for large $m$ all groups have a smaller MC error
estimate than in the small $m$
region. Here the numerator of $\chi$ tends to be larger and especially
the denominator becomes very large, both resulting in a slightly better
error estimate for the MC results.

Opposed to MC results, the symmetrized quadrature rules give error estimates approximately
at machine precision up to very small $m$ values, see Figure \ref{pic:sym0d_chiralCondensate_error_1024bitPrecision}.
These numerical results show that the symmetrized quadrature rules
give significant results in the sign-problem region in practice, where MC
simulations have error estimates of order one.


\subsection{Reducing high-dimensional integrals to nested
  one-dimensional integrals}
\label{ssec:RNI}
The previous section shows efficient quadrature rules for physical one-dimensional
integration to avoid the sign-problem. Most physical models have more than
one integration variable. In general, it is not
straightforward to find an efficient quadrature rule, and usually
restricted Monte Carlo methods are applied to high-dimensional integrals. As a first alternative, we investigated the recursive numerical integration(RNI) method. This method reduces the
$d$-dimensional integral
\begin{align}
  I(f) = \int_{D^d} f[\varphi] \,\dif \varphi
  \label{equ:symxd_int}
\end{align}
with $\dif \varphi = \prod_{i=1}^d \dif \varphi_i$ and $D=[0,2\pi)$ into many recursive
one-dimensional integrals, and can be applied for several physical models of interest. 

This is done by utilizing the typical structure of the integrand
$f[\varphi]$. This section focuses on the RNI
method and how to find an efficient quadrature rule for a
high-dimensional integral. It does not discuss the sign-problem which
is investigated further in section \ref{ssec:sym1}. More specifically,
this section describes the two
steps to create an efficient quadrature rules $Q(f)$ for the integral
$I(f)$ in \eqref{equ:symxd_int}:
\begin{enumerate}
  \item[RNI 1.] Use the structure of the integrand of the high-dimensional integral to
    rewrite it into recursive
    one-dimensional integrals.
    \item[RNI 2.] Choose an efficient quadrature rule to compute each
      one-dimensional integral numerically. Recursively doing this
      results in the full quadrature rule $Q(f)$.
\end{enumerate}
Finally, this section shows results of applying the method to a physical model called the
topological osciallator. A more detailed explanation of the method and
the results can
be found in \cite{Ammon:2015mra, Ammon:2016dfi}.

\subsubsection{RNI 1. Using the structure of the integrand}
Many models in one physical dimensional have integrands with the structure
\begin{align}
    f[\varphi] = \prod_{i=1}^d f_i(\varphi_{i+1}, \varphi_i),
  \label{equ:RNI_decompositionOfIntegrand}
\end{align}
with periodic boundary conditions $\varphi_{d+1} = \varphi_1$. These
models have only next-neighbor couplings.

The integral of \eqref{equ:RNI_decompositionOfIntegrand} can be
rewritten using recursive integration as described in \cite{Genz86,
  Hayter06}: Because
of next-neighbor couplings, each variable $\varphi_i$ appears only
twice in $f[\varphi]$, in $f_i$ and
$f_{i-1}$, and therefore the integral can be written as $d$ nested
one-variable integrals $I_i$,
\begin{align}
  &I(f) = \int_{D} ... \int_{D} \prod_{i=1}^d
    f_i(\varphi_i, \varphi_{i+1}) \, \dif\varphi_d \cdots \dif\varphi_1
    \label{equ:RNI_recursiveIntegral}\\
  &\resizebox{.9\hsize}{!}{$\displaystyle{
   =\underbrace{\int_{D}\Bigg( ...
    \underbrace{\bigg(\int_{D} f_{d-2}(\varphi_{d-2},
  \varphi_{d-1}) \cdot \underbrace{\bigg(\int_{D} f_{d-1}(\varphi_{d-1}, \varphi_d)
  \cdot f_d(\varphi_d, \varphi_{d+1}) \,\dif \varphi_d \bigg)}_{I_d}
    \, \dif \varphi_{d-1} \bigg)}_{I_{d-1}} \cdots\Bigg) \,\dif \varphi_1}_{I_1}
    }$}.
    \nonumber
\end{align}
This full integral can be computed recursively: $I_d$ integrates out
$\varphi_d$ first, then $I_{d-1}$ integrates out $\varphi_{d-1}$ and so
on until finally $I_1 = I(f)$ integrates out $\varphi_1$. 

To avoid under- and overflow of the
single quadrature rule results, we actually used quadrature rules to
approximate $I_i^* = \frac{1}{c_i} I_i$ with
$c_i > 0$ chosen adaptively. Then, the final integral is computed via
$I = \left(\prod_{i=1}^d c_i\right) I^*$. For brevity,
the method is described in the following without this trick.

Each integral is approximated by using an $N_{\text{quad}}$-point quadrature rule.
The first integrand in \eqref{equ:RNI_recursiveIntegral} (last from the right) 
depends on three variables $\varphi_{d-1}$, $\varphi_d$ and
$\varphi_{d+1}$. The variable $\varphi_d$ is integrated out, therefore
the quadrature rule $Q_d(f_{d-1} \cdot f_d) \equiv Q_d$ of $I_d$ depends
on two variables,
\begin{align}
  Q_d(\varphi_{d-1}, \varphi_{d+1}) =
  \sum_{\gamma=1}^{N_{\text{quad}}} w_\gamma \,\, f_{d-1}(\varphi_{d-1},t_\gamma) \,\, f_d(t_\gamma,
           \varphi_{d+1}),
  \label{equ:RNI_quadratureRule_Id}
\end{align}
with sampling points $t$ and weights $w$.
The next integral $I_{d-1}$ is approximated by the quadrature rule
\begin{align}
  Q_{d-1}(\varphi_{d-2}, \varphi_{d+1}) =
  \sum_{\gamma=1}^{N_{\text{quad}}} w_\gamma \,\,
  f_{d-2}(\varphi_{d-2},t_\gamma) \,\, Q_d(t_\gamma, \varphi_{d+1}),
  \label{equ:RNI_quadratureRuleId-1}
\end{align}
and includes the quadrature rule $Q_d$ given in \eqref{equ:RNI_quadratureRule_Id}.
The quadrature rules $Q_{d-2}$, ..., $Q_1$ are created analogically to \eqref{equ:RNI_quadratureRuleId-1}.
Using the same sampling points $w_\gamma$ and weights $t_\gamma$, $\gamma \in \{1,...,N_{\text{quad}}\}$
in all quadrature rules  $Q_i$ results in the full quadrature rule for \eqref{equ:RNI_recursiveIntegral},
\begin{align}
  Q = Q_1 = \sum_{\gamma=1}^{N_{\text{quad}}} w_\gamma Q_2(t_\gamma, t_\gamma) = \trlong{ \prod_{i=1}^d
        \left(M_i \cdot
  \diag{(w_1,\dots,w_{N_{\text{quad}}})}\right)},
  \label{equ:RNI_quadratureRule_I1}
\end{align}
with $M_i$ beeing an $N_{\text{quad}} \times N_{\text{quad}}$ matrix
with entries $(M_i)_{\alpha \beta} = f_i(t_\alpha, t_\beta)$.

\subsubsection{RNI 2. Choosing an efficient quadrature rule}
We used the Gaussian-Legendre $N_{\text{quad}}$-point quadrature
rule, see \cite{stoer2013introduction} to define the sampling points
$t$ and weights $w$. For this rule, the error scales asymptotically
(for large $N_{\text{quad}}$) as
$\sigma \sim \order\left(\frac{1}{(2N_{\text{quad}})!}\right)$ (For Legendre polynomials the correct asymptotic error
  scaling is $\frac{(N_{\text{quad}}!)^4}{((2N_{\text{quad}})!)^3}$ \cite{kahaner1989numerical}
  which is slightly improved over $\frac{1}{(2N_{\text{quad}})!}$.). The Stirling formula
($N_{\text{quad}}! \approx \sqrt{2\pi N_{\text{quad}}} \left(\frac{N_{\text{quad}}}{e}\right)^{N_{\text{quad}}}$ asymptotically)
approximates the factorial to give
\begin{align}
 \sigma \sim \order\left( \exp(-2N_{\text{quad}}\ln
N_{\text{quad}}) \frac{1}{\sqrt{N_{\text{quad}}}} \right)
  \label{equ:RNI_errorScaling}
\end{align}
asymptotically. This is a huge improvement over the MC error scaling $1 / \sqrt{N_{\MC}}$.

\subsubsection{Application to the topological oscillator}
\label{sssec:topologicalOsciallator}
We applied the RNI method to the topological
oscillator \cite{Bietenholz:2010xg}, also called quantum rotor, which is a simple, physically
one-dimensional model that has non-trivial characteristics which are also
present in more complex models. It has $d$ variables $\phi_i \in [0,
2\pi)$, a length scale $T$ and a coupling constant $c$. We investigated the topological charge
susceptibility of this model,
\begin{align}
  \chi_{\text{top}} &= \frac{\int O[\varphi] B[\varphi] \, \dif \varphi}
                      {\int B[\varphi] \,\dif\varphi},
                      \label{equ:RNI_topologicalChargeSusceptibility}
\end{align}
with Boltzmann-weight
\begin{align}
B[\varphi] &= \exp{\left( -c \sum_{i=1}^d \left(1 -
                    \cos(\varphi_{i+1} -
  \varphi_i) \right) \right)},
\end{align}
and a squared topological charge
\begin{align}
  O[\varphi] &= \frac{1}{T}\left(\frac{1}{2\pi} \sum_{i=1}^d (\varphi_{i+1} - \varphi_i) \mod
  2\pi\right)^2.
\end{align}

With RNI, we computed both numerator and denominator of
$\chi_{\text{top}}$ separately, both
differing in the factorization
\eqref{equ:RNI_decompositionOfIntegrand} of their integrands.
Straightforwardly, the denominator integrand consists out of local exponential
factors. The numerator consists out of summands with varying
factorization schemes, each of these summands is computed separately with RNI
and they are presented in more detail in \cite{Volmer2018New}, section 5.2.
We estimated the error of $\chi_{\text{top}}$ by choosing a large
number of samples $N_{\text{quad}}^g$ in
\eqref{equ:RNI_quadratureRule_Id}, \eqref{equ:RNI_quadratureRuleId-1}
and similar ones
for which we assumed that $\chi_{\text{top}}(N_{\text{quad}}^g)$ has
converged to the actual value and
computed the difference of $\chi_{\text{top}}(N_{\text{quad}})$ for
$N_{\text{quad}} < N_{\text{quad}}^g$ to this value,
\begin{align}
  \Delta \chi_{\text{top}}(N_{\text{quad}}) =
  |\chi_{\text{top}}(N_{\text{quad}}) -
  \chi_{\text{top}}(N_{\text{quad}}^g)|.
  \label{equ:RNI_truncationError}
\end{align}
We tested beforehand that this truncation error behaves exponentially
for large $N_{\text{quad}}$ in practice, as expected from \eqref{equ:RNI_errorScaling},
\cite{Ammon:2015mra}.

\begin{figure}
  \includegraphics[width=1.\textwidth]{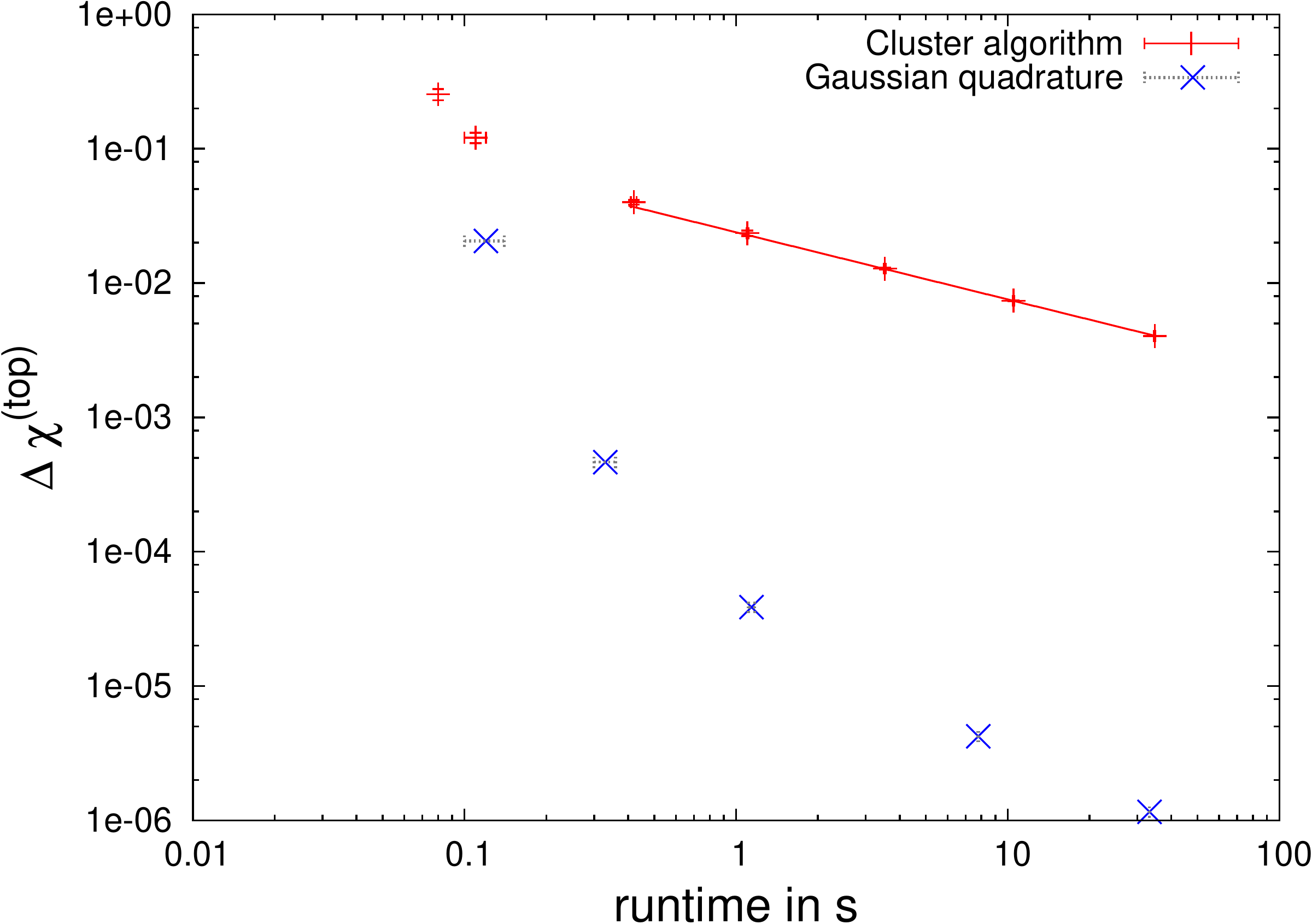}
  \caption{The runtime to arrive at a given
    error estimate is orders of magnitudes smaller when using the RNI
    method with Gauss-Legendre points than using the Cluster MCMC algorithm.}
  \label{pic:RNI_susceptibilityError_clusterAndRNI}
\end{figure}

We compared the results of the RNI method with results using the
Cluster algorithm \cite{Wolff:1988uh}, which we found is an optimal
MCMC method for the application to the topological
oscillator \cite{Ammon:2015mra}. 
Due to the exponential error scaling of the Gauss-Legendre rule, the new method
advances MCMC for large enough $N_{\text{quad}}$. We found that the RNI method
is also advantageous for lower $N_{\text{quad}}$-values:
our simulations showed that the RNI method needs orders of magnitude
less runtime than the Cluster algorithm to result in a specified error
estimate on an observable, compare Figure
\ref{pic:RNI_susceptibilityError_clusterAndRNI} for $c = 2.5$, $T = 20$, $d = 200$.
The Cluster algorithm measurements resulted in an error estimate that
decreases proportional to $t^{-1/2}$ for runtime $t$, consistent with
the typical MC error scaling \cite{probTheory}. We used between $10^2$
and $10^6$ sampling points here. The RNI method, using between 10 and
300 sampling points with $N_{\text{quad}}^g=400$, resulted in orders
of magnitude smaller errors. The exponential error scaling in
\eqref{equ:RNI_errorScaling} is not visible here, the asymptotic
regime of the method is not yet reached with the used numbers of
sampling points.

All in all, the RNI method results in orders of magnitude smaller
errors than the Cluster algorithm for a fixed runtime or equivalently,
the RNI method needs orders of magnitude less runtime than the Cluster
algorithm to arrive at a fixed error estimate, even for a number of
sampling points where the RNI error does not yet scale exponentially.

\subsection{Avoiding the sign-problem in high-dimensional integrals}
\label{ssec:sym1}
Section \ref{ssec:sym0} shows that the sign-problem can be avoided for
one-dimensional integrals using symmetric quadrature rules. 
But what about the sign-problem for
high-dimensional integrals? A quadrature rule for high-dimensional
integrals over compact groups,
\begin{align}
  I(f) = \int_{G^d} f[U] \, \dif U,
  \label{equ:combined_int}
\end{align}
with $\dif U = \prod_{i=1}^d \dif U_i$ is needed that also avoids the sign-problem. We combined both already
presented methods, the
symmetric quadrature rules in section \ref{ssec:sym0} and the RNI in
section \ref{ssec:RNI} to
find an efficient quadrature rule $Q(f)$ for $I(f)$ in
\eqref{equ:combined_int}. An alternative attempt to generalize the
symmetrized quadrature rules to high-dimensional integrals is
discussed in \cite{10.1007/978-3-319-75996-8_15}.

\subsubsection{Combining recursive numerical integration and symmetric
  quadrature rules}
RNI can be used to transform the high-dimensional integral $I(f)$ in
\eqref{equ:combined_int} into one-dimensional integrals. These
one-dimensional integrals can be approximated recursively, using the
symmetric quadrature rules. In the following, these steps are described
in more detail:
\begin{enumerate}
\item[RNI 1.] Find the structure, i.e. all $f_i$, of the integrand
  \begin{align}
    f[U] = \prod_{i=1}^d f_i(U_{i+1}, U_i),
  \end{align}
  to be able to write the full integral as nested one-dimensional
  integrals, similar to \eqref{equ:RNI_recursiveIntegral}.
\item[RNI 2.] Apply symmetric quadrature rules to each one-dimensional
  integration over $U_i$. Here is an example how to do this for the
  innermost integral $I_d$, integrating over $U_d$:
  \begin{enumerate}
  \item[Sym 1.] Rewrite the integral over $U_d$ into an integral over
    the products of spheres as done in
    \eqref{equ:sym01_integralOverSpheresToCompactGroups}.
  \item[Sym 2.] Approximate each  iterated integral $I_{d}(g)$ by a product rule of quadratures over spheres parametrising the group $U_d$ to be integrated. Note that the group $U_d$ is parametrised at most as the product of $S^1$, $S^3 $ and $S^5$.
\end{enumerate}
\end{enumerate}
\subsubsection{Application to topological oscillator with
  sign-problem}
We applied this combined method again to the topological oscillator
discussed in \ref{sssec:topologicalOsciallator}. This time we
transformed the variables $\varphi_i$ to new variables $U_j =
\e{i\varphi_j} \in \Ugroup{1}$. Additionally, we added a sign-problem to the model by
using an additional factor $\prod_{j=1}^d U_j ^{-\theta}$ in the Boltzmann-weight,
\begin{align}
  B[U] &= \exp{\left(- c \sum_{i=1}^{d} \Re(1 - U_{i+1} U_i^*)
              \right)} \cdot \prod_{j=1}^d U_j ^{-\theta},
\end{align}
with a new parameter $\theta \in \R$. If this parameter is larger than zero,
the sign-problem arises and is most severe for $\theta = \pi$.

In this model we computed the plaquette,
\begin{align}
  plaquette &= \frac{\int O[U] B[U] \, \dif U}
                      {\int B[U] \,\dif U},
              \label{equ:combined_plaquette}
\end{align}
with 
\begin{align}
  O[U] = \frac{1}{d} \Re\left( \sum_{i=1}^{d} U_{i+1} U_i^*\right).
  \label{equ:sym1d_linkCorrelation}
\end{align}

For the combined method, we computed both numerator and denominator of
\eqref{equ:combined_plaquette} separately and divided the 
values. We used a truncation error, similar to the one given in \eqref{equ:RNI_truncationError}.
We compared the method with a standard MC method as used
in \ref{sssec:1dQCD}. The MC error is computed via the standard deviation.

\begin{figure}
  \includegraphics[width=.9\textwidth]{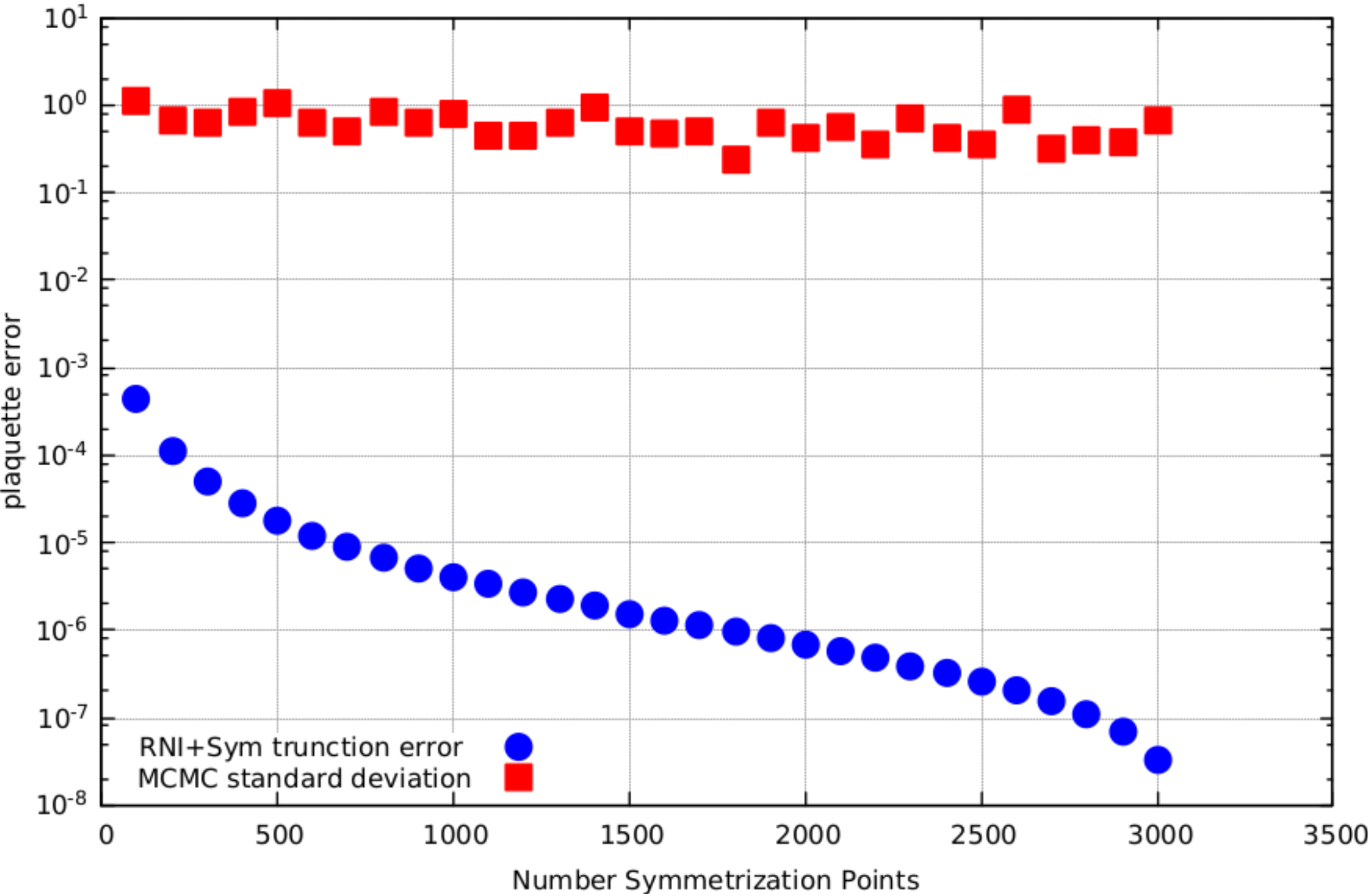}
  \caption{The combined method avoids the sign-problem that exists
    when using the MC method.}
  \label{pic:combined_result}
\end{figure}

For $\theta=\pi$ we found that the combined method avoids the sign-problem that is
visible with the MC computation, compare Figure
\ref{pic:combined_result}. It gives orders of magnitude smaller errors that shrink
the more symmetrization points are used.
Therefore the combination of RNI and symmetric quadrature rules is
suitable to avoid
the sign-problem for high-dimensional integration.

\section{Conclusion}

In this contribution we have demonstrated that through symmetric
quadrature rules
exact symmetrization
and recursive numerical 
integration techniques problems in high energy physics can be solved 
which constitute a major, if not unsurmountable obstacle for standard 
Markov chain Monte Carlo methods. 
The examples we have considered here 
invole only a time lattice and are hence 0+1-dimensional in space-time, where as 
real physical problem include spatials dimensions of up to 3. We are presently 
investigating whether the methods we have presented here can be extended 
to higher, i.e. including also spacial, dimensions. While for the recursive numerical integration 
technique we have first results which are promising, for the full symmetrization 
method we were so far not successful. 

Also combining the symmetrized quadrature rules with MC methods did not 
lead to a practically feasible method in higher dimensions. However, we are 
following a path to combine Quasi Monte Carlo, recursive numerical 
integration and a full symmetrization to overcome this problem and hope to 
report about these attempts in the future. 


%
\bibliographystyle{spmpsci}
%


\end{document}